\documentstyle[epsf,aps]{revtex}
\newcommand{\mafigura}[4]{
  \begin{figure}[hbtp]
    \begin{center}
      \epsfxsize=#1 \leavevmode \epsffile{#2}
    \end{center}
    \caption{#3}
    \label{#4}
  \end{figure} }
\newcommand{\beq}{\begin{equation}}
\newcommand{\eeq}{\end{equation}}
\newcommand{\bea}{\begin{eqnarray}}
\newcommand{\eea}{\end{eqnarray}}

\begin{document}
\preprint{TTP98-07}

\title{Charge asymmetry in hadroproduction of heavy quarks}

\draft

\author{J.H. K\"uhn and G. Rodrigo}
\address{Institut f\"ur Theoretische Teilchenphysik,
Universit\"at Karlsruhe\\ D-76128 Karlsruhe, Germany}

%\date{\today}
\date{February 12, 1998}

\maketitle
 
\begin{abstract}
A sizeable difference in the differential production cross section
of top and antitop quarks, respectively, is predicted for 
hadronically produced heavy quarks.
It is of order $\alpha_s$ and arises from the 
interference between charge odd and even amplitudes respectively.
For the TEVATRON it amounts up to 15\% for the differential 
distribution in suitable chosen kinematical regions.
The resulting integrated forward-backward asymmetry of 
4--5\% could be measured in the next round of experiments.
At the LHC the asymmetry can be studied by selecting 
appropriately chosen kinematical regions.
\end{abstract}


%\narrowtext

%%%%%%%%%%%%%%%%%%%%%%%%%%%%%%%%%%%%%%%%%%%%%%%%%%%%%%%%%%%%
%%%%%%%%%%%%%%%%%%%%%%%%%%%%%%%%%%%%%%%%%%%%%%%%%%%%%%%%%%%%

Top quark production at hadron colliders has become one of the
central issues of theoretical~\cite{Catani:1997rn}
and experimental~\cite{Tipton:1996}
research. The investigation and understanding of the production 
mechanism is crucial for the determination of the top quark 
couplings, its mass and the search for new physics involving the top
system. 
A lot of effort has been invested in the prediction of the total 
cross section and, more recently, of inclusive transverse
momentum distributions~\cite{Catani:1997rn}. 

In this work we will point to a different aspect of the hadronic
production process, which can be studied with a 
fairly modest sample of quarks.
Top quarks produced through light quark-antiquark annihilation 
will exhibit a sizeable charge asymmetry -- an excess of 
top versus antitop quarks in specific kinematic regions --
induced through the interference of the final-state
with initial-state radiation (Fig.~\ref{fig:diagrams} a, b)
and the interference of the box with the lowest-order-diagram
(Fig.~\ref{fig:diagrams} c, d).
The asymmetry is thus of order $\alpha_s$ relative to the 
dominant production process.
In suitable chosen kinematical regions it reaches 
up to 15\%, the integrated forward-backward asymmetry amounts  
to 4--5\%.
Top quarks are tagged through their decay $t\rightarrow b \; W^+$
and can thus be distinguished experimentally from antitop 
quarks through the sign of the lepton in the 
semileptonic mode and eventually also through the $b$-tag. 
A sample of hundred to two hundred tagged top quarks 
should in fact be sufficient for a first indication 
of the effect.

%%%%%%%%%%%%%%%
\mafigura{8 cm}{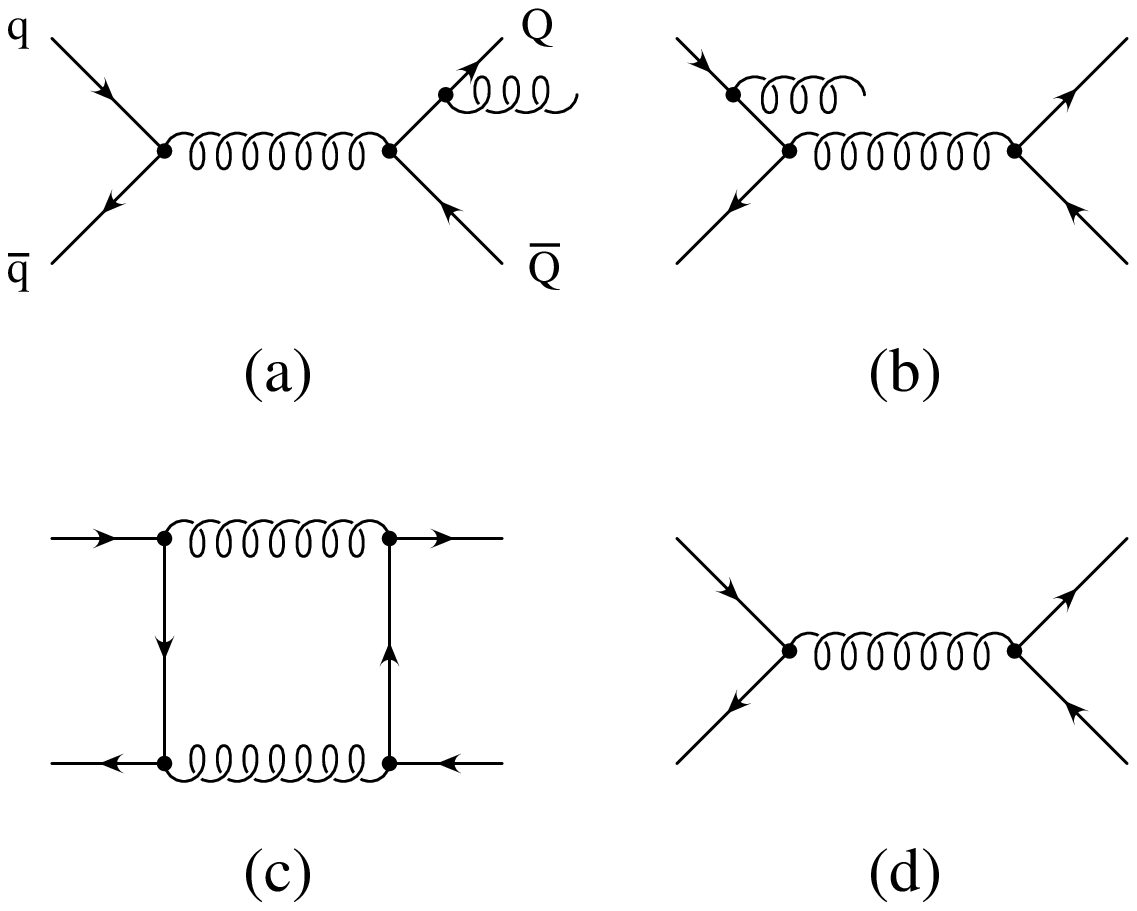}{Origin of the QCD charge
asymmetry in hadroproduction of heavy quarks:
interference of final-state (a) with initial-state (b) gluon bremsstrahlung
plus interference of the box (c) with the Born diagram (d).}
{fig:diagrams}
%%%%%%%%%%%%%%%

Top production at the TEVATRON is dominated by quark-antiquark
annihilation, hence the charge asymmetry will be reflected
not only in the partonic rest frame but also in the center 
of mass system of proton and antiproton. The situation 
is more intricate for proton-proton collisions at the LHC, 
where no preferred direction is at hand in the laboratory frame. 
Nevertheless it is also in this case possible to pick 
kinematical configurations which allow the study of the charge
asymmetry.

The charge asymmetry has also been investigated in~\cite{Halzen:1987xd}
for a top mass of $45$ GeV. There, however, only the contribution 
from real gluon emission was considered requiring the 
introduction of a physical cutoff on the gluon energy and 
rapidity to avoid infrared and collinear singularities.
Experimentally, however, only inclusive top-antitop production has been 
studied to date, and the separation of an additional soft 
gluon will in general be difficult. 
In this work, we will therefore include virtual corrections and 
consider inclusive distributions only.
We will see below, that the sign of the asymmetry for 
inclusive production is opposite to the one 
given for the $t\bar{t}g$ process in~\cite{Halzen:1987xd}.
The charge asymmetry of heavy flavour production in quark-antiquark
annihilation to bottom quarks was also discussed
in~\cite{Ellis:1986ba,Nason:1989zy,Beenakker:1991ma} where its contribution 
to the forward-backward asymmetry in proton-antiproton collisions
was shown to be very small.
In addition there is also a slight difference between the 
distribution of top and antitop quarks in the reaction 
$g q \to t \bar{t}q$. At the 
TEVATRON its contributions is bellow $10^{-4}$. 
(This effect should not be confused with the large asymmetry in the top
quarks' angular or rapidity distribution in this reaction which 
is a trivial consequence of the asymmetric partonic initial 
state and vanishes after summing over the incoming parton beams.)

In a first step the charge asymmetry will be evaluated at the
partonic level for the quark-antiquark induced reaction. 
The calculation proceeds in analogy to the corresponding 
QED process~\cite{Berends:1973,Berends:1983dy}.
The interference terms corresponding to real emission
(Fig.~\ref{fig:diagrams} a$*$b) and virtual radiation 
(Fig.~\ref{fig:diagrams} c$*$d) are evaluated separately 
with an appropriate infrared regulator.
Soft radiation up to a cutoff $E_{cut}^g$ is then combined 
with the virtual correction leaving the hard radiation 
with $E^g>E_{cut}^g$ which can be evaluated numerically. 
The asymmetric part does not exhibit a light quark 
mass singularity, whence $m_q$ can be set to zero throughout; in 
other words, no collinear singularities arise.
The virtual plus soft radiation on one hand and the real 
hard radiation on the other contribute 
with opposite signs, with the former always larger than 
the later which explains the difference in sign 
between our result and~\cite{Halzen:1987xd}.

%%%%%%%%%%%%%%%
\mafigura{6.8 cm}{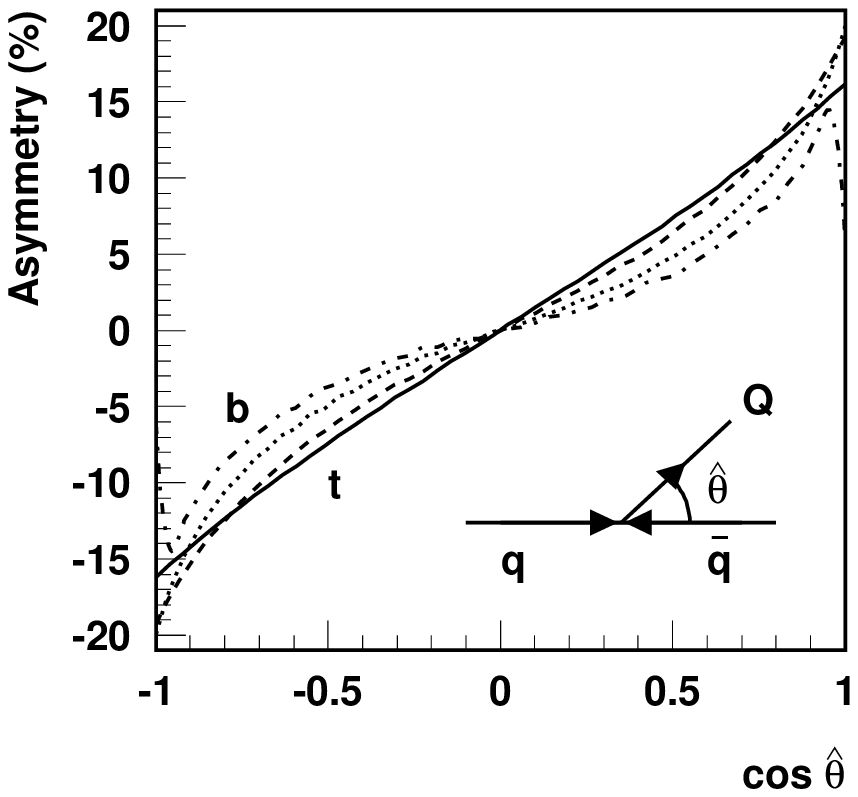}{Differential charge asymmetry 
in top quark pair production for fixed partonic center of
mass energy $\sqrt{\hat{s}}=400$ GeV (solid),
$600$ GeV (dashed) and $1$ TeV (dotted).
We also plot the differential asymmetry for a b-quark 
with $\sqrt{\hat{s}}=400$ GeV (dashed-dotted).}
{fig:sfix}
%%%%%%%%%%%%%%%
%%%%%%%%%%%%%%%
\mafigura{6.8 cm}{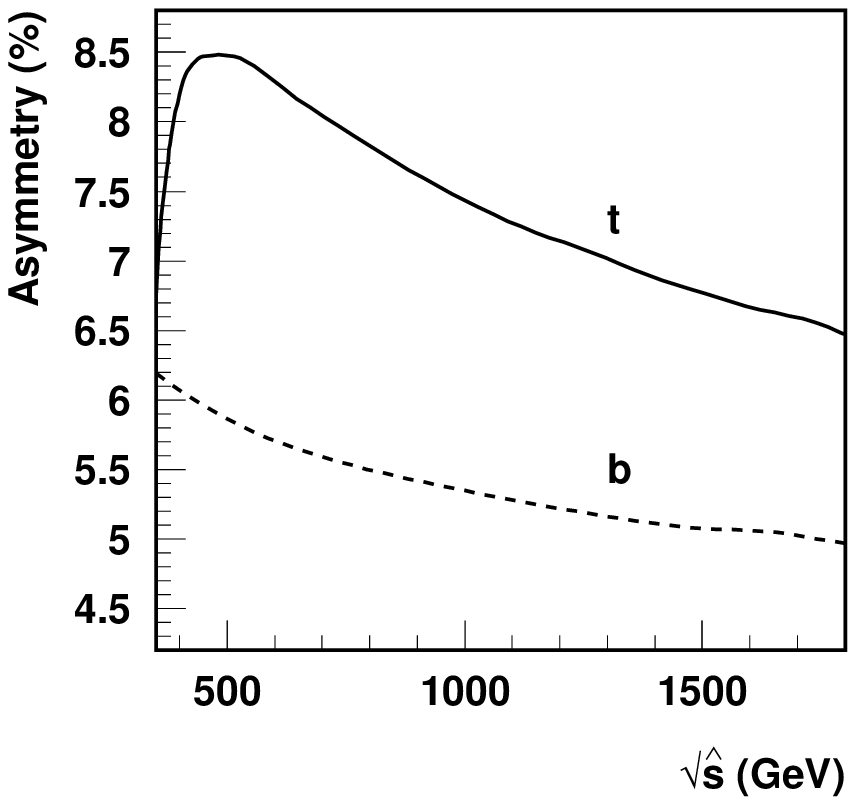}{Integrated charge
asymmetry as a function of the partonic center or mass energy
for top and bottom quark pair production.}
{fig:sdistr}
%%%%%%%%%%%%%%%

The QCD asymmetry is related to the 
corresponding QED asymmetry through  the replacement of 
$\alpha_{QED} Q Q'$ by the factor 
$\frac{1}{2}\alpha_s (d_{abc}/4)^2 = \alpha_s \cdot 5/12$.
Let us define the differential asymmetry through
\beq
\hat{A}(\cos \hat{\theta}) = 
\frac{N_t(\cos \hat{\theta})-N_{\bar{t}}(\cos \hat{\theta})}
     {N_t(\cos \hat{\theta})+N_{\bar{t}}(\cos \hat{\theta})}~,
\eeq
where $\hat{\theta}$ denotes the top quark production angle in 
the $q \bar{q}$ restframe and $N(\cos \hat{\theta}) =
d\sigma/d\Omega (\cos \hat{\theta})$.
Since $N_{\bar{t}}(\cos \hat{\theta}) = N_t(-\cos \hat{\theta})$
as a consequence of charge conjugation symmetry, 
$\hat{A}(\cos \hat{\theta})$ can also be interpreted as
a forward-backward asymmetry of top quarks.
In Fig~\ref{fig:sfix}, $\hat{A}(\cos \hat{\theta})$
is displayed for $\sqrt{\hat{s}}=400$~GeV, $600$~GeV and $1$~TeV
for $M_t = 175$~GeV.
For completeness we also display the result for $b \bar{b}$
production at $\sqrt{\hat{s}}=400$~GeV
with $M_b=4.6$~GeV.
The strong coupling constant is evaluated at the scale
$\mu=\sqrt{\hat{s}}/2$ from $\alpha_s(M_Z) = 0.118$.

The integrated charge asymmetry
\beq
\bar{\hat{A}} = 
\frac{N_t(\cos \hat{\theta} \geq 0)
    - N_{\bar{t}}(\cos \hat{\theta} \geq 0)}
     {N_t(\cos \hat{\theta} \geq 0)
    + N_{\bar{t}}(\cos \hat{\theta} \geq 0)}~,
\eeq
is shown in Fig~\ref{fig:sdistr} as a function of $\sqrt{\hat{s}}$.
With a typical value around $6-8.5 \%$  it should be well 
accessible in the next run of the TEVATRON.

%%%%%%%%%%%%%%%
\mafigura{7.5 cm}{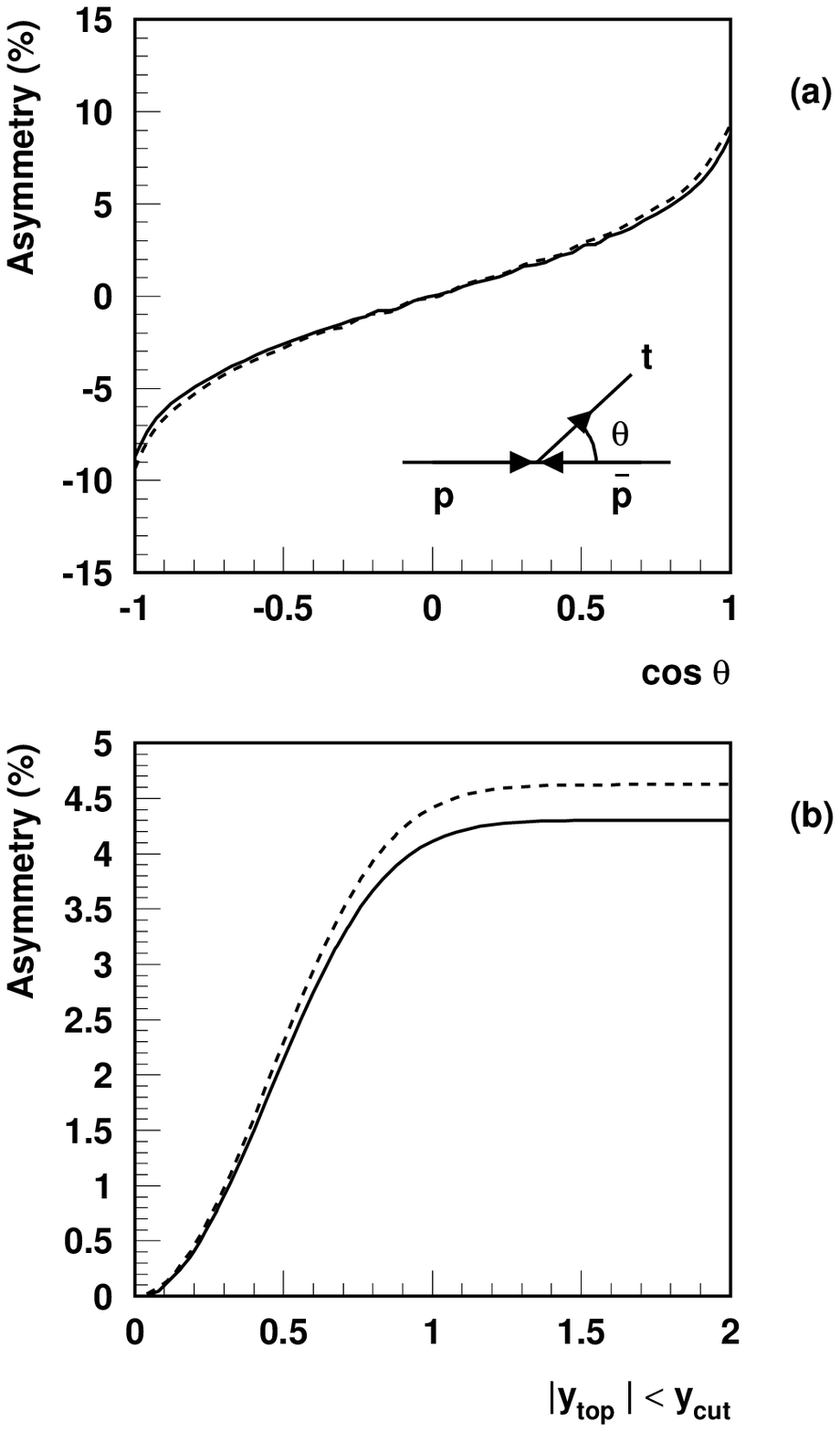}{a) Differential charge asymmetry
in the proton-antiproton restframe using the
MRS96-1 structure function. We consider also 
two different choices of the factorization scale: $\mu=\sqrt{\hat{s}}$
(solid) and $\mu=\sqrt{\hat{s}}/2$ (dashed).
b) Integrated asymmetry for (anti-)top quarks with rapidities
less than $y_{cut}$.}
{fig:PDFdistr}
%%%%%%%%%%%%%%%

The asymmetry can in principle be studied experimentally in 
the partonic restframe, as a function of $\hat{s}$, by 
measuring the invariant mass of the $t \bar{t}$ system 
plus an eventually radiated gluon. It is, however, also instructive
to study the asymmetry in the laboratory frame by folding 
the angular distribution with the structure
functions~\cite{Martin:1996as}.
The differential asymmetry is displayed in Fig.~\ref{fig:PDFdistr}a,
where $q \bar{q}$ and $g g$ initiated processes are 
included in the denominator. For the total 
charge asymmetry we predict 
\beq
\bar{A} = 
\frac{N_t(\cos \theta \geq 0)
    - N_{\bar{t}}(\cos \theta \geq 0)}
     {N_t(\cos \theta \geq 0)
    + N_{\bar{t}}(\cos \theta \geq 0)} = 4.3 - 4.6 \%~,
\eeq
where different choices of the structure function and 
different choices of the factorization scale, $\mu = \sqrt{\hat{s}}$
and $\mu = \sqrt{\hat{s}}/2$, have been considered.

%%%%%%%%%%%%%%%
%\mafigura{13 cm}{relamount_old.ps}{
\mafigura{8 cm}{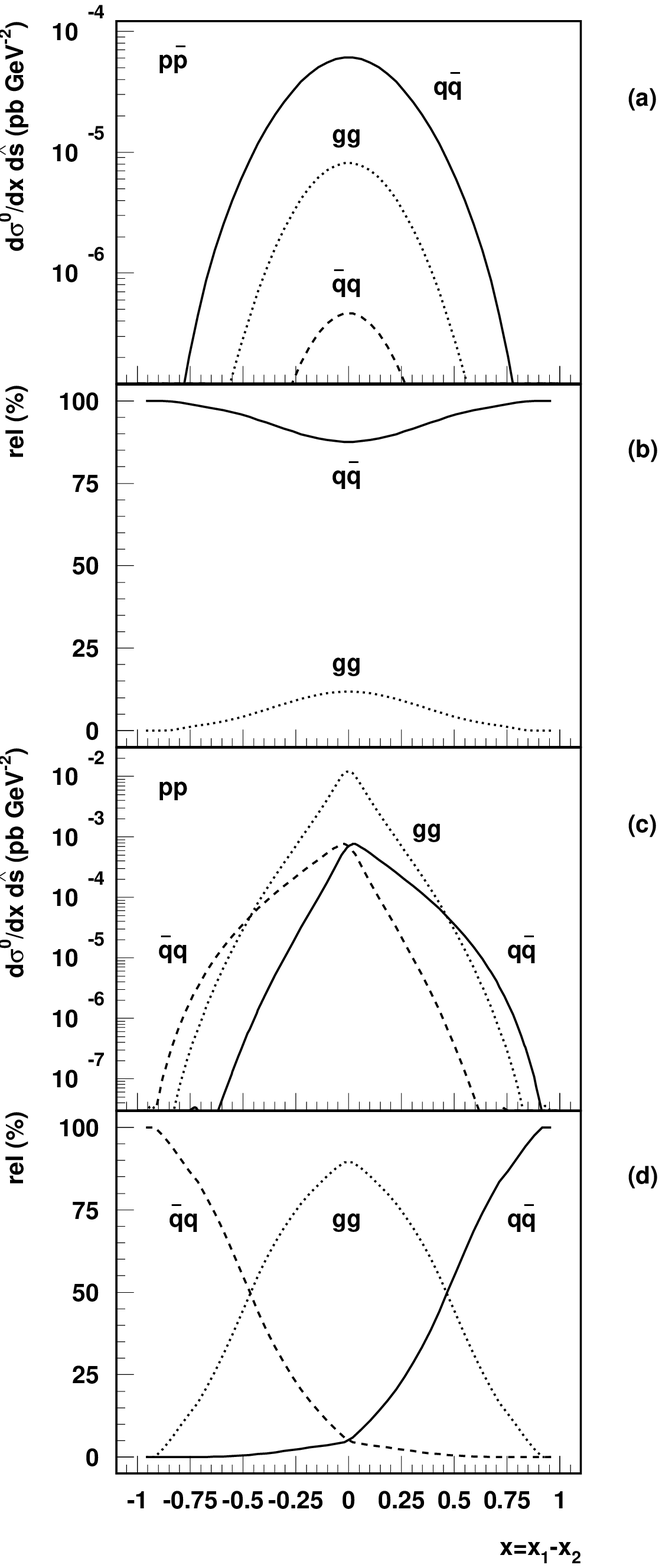}{
Differential cross sections (Fig.\ a, c) and
relative amount (Fig.\ b, d) of quark-antiquark, antiquark-quark 
and gluon-gluon initiated processes as functions of 
$x_1-x_2=2 P_3(t \bar{t} g)/\sqrt{s}$ in lowest order,
for $\sqrt{s}=1.8$~TeV in proton-antiproton (Fig. a, b) 
and $\sqrt{s}=14$~TeV in proton-proton (Fig. c, d) collisions with
$\sqrt{\hat{s}}=400$~GeV in both cases.}
{fig:relamount}
%%%%%%%%%%%%%%%

In principle one might expect that cuts on the top quark or its
decay products at large rapidities could affect the asymmetry.
In Fig~\ref{fig:PDFdistr}b we thus present the asymmetry for the 
restricted range $\mid y_{top}\mid < y_{cut}$ as a function 
of $y_{cut}$. It approaches its maximal value already for 
$y_{cut}=1$, indicating that also cuts on the top decay products 
$W$ and $b$ jets with rapidities, say, larger than 2 will not lead
to a significant reduction of the asymmetry.
We would also like to mention that event generators which do not 
include the full NLO matrix
elements~\cite{Marchesini:1991ch,Sjostrand:1994yb} cannot 
predict the asymmetry.

Top-antitop production in proton-proton collisions at the 
LHC is, as a consequence of charge conjugation symmetry,
charge symmetric if the laboratory frame 
is chosen as the reference system. However, by selecting
the invariant mass of the $t \bar{t} (+g)$ system and its
longitudinal momentum appropriately, one can easily constrain
the parton momenta such that a preferred direction is 
generated for quark-antiquark reactions.
This last point is illustrated in Fig~\ref{fig:relamount}
where we present the relative amount of quark-antiquark,
antiquark-quark and gluon-gluon initiated processes as functions of 
$x_1-x_2=2 P_3(t \bar{t} g)/\sqrt{s}$ in lowest order,
for $\sqrt{s}=14$~TeV and $\sqrt{\hat{s}}=400$~GeV
as characteristical example.
A detailed study of this situation will be presented 
elsewhere~\cite{KR}.

The box diagram, Fig.~\ref{fig:diagrams} c, can also give 
rise to $t \bar{t}$ in a colour singlet configuration, 
which in turn interferes with $t \bar{t}$ production through the 
photon or Z.
A similar consideration applies to interference between 
initial and final state radiation.
The resulting asymmetry is obtained from the QCD asymmetry through the 
following replacement 
\beq
\frac{\alpha_s}{2} \left( \frac{d_{abc}}{4} \right)^2
\rightarrow 
\alpha_{QED} \left( Q_t Q_q + 
\frac{(1-\frac{\displaystyle 8}{\displaystyle 3} s^2_W)(2 I_q-4 Q_q s^2_W)}
{16 s^2_W c^2_W}  
\frac{1}{1-\frac{\displaystyle M_Z^2}{\displaystyle \hat{s}}} \right) 
\eeq
which amounts to an increase of the asymmetry by 
typically a factor $1.04$ and is thus smaller than 
uncalculated higher order corrections.

To summarize: the charge asymmetry can be used as an important
tool to investigate the production dynamics. For the 
TEVATRON it amounts to roughly $4-5\%$ and can therefore 
be studied with a sample of several hundred $t \bar{t}$
pairs expected for the next run. 
The asymmetry can also be studied at the LHC if one selects 
appropriate kinematic configurations.

\vspace{.5cm}

We would like to acknowledge useful discussions with R.K.~Ellis,
T.~Sj\"ostrand and M.~Seymour.
Work supported by BMBF under Contract 057KA92P 
and DFG under Contract Ku 502/8-1.

%\bibliographystyle{utphys}
%\bibliography{asymetry}

\providecommand{\href}[2]{#2}\begingroup\raggedright\endgroup

\end{document}